\documentclass[10pt,preprintnumbers,twocolumn,amsmath,amssymb,nofootinbib,superscriptaddress]{revtex4-1}
\usepackage[latin1]{inputenc}
\usepackage{slashed}
\usepackage{amsmath}
\usepackage{textcomp}
\usepackage{amssymb}
\usepackage{amsfonts}
\usepackage{indentfirst}
\usepackage{color}
\usepackage{hyperref}
\usepackage{subcaption}

\newcommand{\be}{\begin{equation}}
\newcommand{\ee}{\end{equation}}
\newcommand{\bea}{\begin{eqnarray}}
\newcommand{\eea}{\end{eqnarray}}

\newcommand{\nn}{\nonumber\\}

\newcommand{\Od}{\mathcal{O}} 
\usepackage{graphicx,graphics}
\graphicspath{{figures/}}

\begin{document}

\hfill{KCL-PH-TH/2022-08}

\title{One-loop tunnelling-induced energetics}

\author{Jean Alexandre and Drew Backhouse}  

\affiliation{Theoretical Particle Physics and Cosmology, King's College London, WC2R 2LS, UK}

\begin{abstract}
Tunnelling between degenerate vacuua is allowed in finite-volume Quantum Field Theory, and features remarkable energetic properties, 
which result from the competition of different dominant configurations in the partition function. 
We derive the one-loop effective potential based on two homogeneous vacuua of the bare theory, and we discuss the resulting Null Energy Condition 
violation in $O(4)$-symmetric Euclidean spacetime, as a result of a non-extensive effective action.
\end{abstract}

\maketitle

\section{Introduction}

Tunnelling is an important aspect of Quantum Mechanics, which is suppressed by a large number of degrees of freedom in Quantum Field Theory (QFT),
where Spontaneous Symmetry Breaking (SSB) plays a fundamental role instead. Strictly speaking though, SSB requires an infinite volume to occur, 
and tunnelling happens in any finite volume, although it can take a huge time to settle in a macroscopic system, in which case SSB is a much better approximation.

In path integral quantisation, the equilibrium state arising from tunnelling can be described when taking into account several saddle points, 
which leads to remarkable energetic features, as a consequence of symmetry restoration and its
relation to convexity \cite{convexity}. One consequence of the competition between different saddle points is a non-extensive 
effective action \cite{AT}, which leads to a violation of the Null Energy Condition (NEC) \cite{NECreview}. NEC violation is known in QFT \cite{NECexamples},
as in the Casimir effect for example, which has been used in a cosmological context to describe the possibility to generate dynamically a spacetime expansion \cite{ZS}.
The non-extensive nature of the effective action has been used in \cite{ACP} to provide a dynamical mechanism for a cosmological bounce \cite{cosmobounce},
without the need for exotic matter or modified gravity. Assuming a contraction of the Universe, the NEC violation mechanism described in \cite{ACP}
switches on when the causal volume has shrunk enough for tunnelling to become significant, consequently leading to a spacetime expansion. 
Tunnelling is then suppressed when the causal volume becomes large enough, and SSB takes over
for the following steps of the cosmological evolution.

In the present article we extend the flat-spacetime study \cite{AT} to one-loop, and we show that quantum corrections do not change the qualitative 
predictions of the "tree-level semi-classical approximation" (ignoring fluctuation factors), regarding the dynamical generation of a non-extensive one-particle-irreducible (1PI) effective action. 
Since this study is done at finite volume, it does not involve any thermodynamical limit, and the famous Maxwell cut which features a flat effective potential cannot be obtained.
It is interesting to note that the Maxwell cut appears in the Wilsonian effective potential \cite{Wilsonian} independently of the volume though, 
but the Wilsonian effective potential is equivalent to the 1PI effective potential for large volume only, and we focus here on the latter potential. 

We comment here on a possible ambiguity between real time and Euclidean time dependence. We are looking at the equilibrium effective action, 
obtained in principle for a large real time. On the other hand, a finite Euclidean time represents the inverse temperature of 
the equilibrium finite-temperature system, and is independent of the typical spacial length involved in the system. 
Tunnelling at finite-temperature is studied in \cite{AP2}, where the Euclidean time can be large, 
therefore allowing (Euclidean) time-dependent saddle points to develop,  
while the space volume is kept finite in order to keep a significant tunnelling rate.
The present work assumes an $O(4)$-symmetric Euclidean spacetime instead, where the length in the "time" direction corresponds to the 
typical time needed for quantum fluctuations to travel though the three-dimensional box in which the scalar field lives, 
and for which the equilibrium is assumed to be reached.

Related to the finite-temperature analogy, we stress that the symmetry restoration mechanism we study here is not related to the 
Kibble-Zurek mechanism \cite{KZ}, which describes high-temperature symmetry restoration in a large three-dimensional volume. Tunnelling-induced
symmetry restoration is valid at any temperature (including zero temperature), therefore in situations where SSB would happen if the volume was infinite.

In Sec.\ref{finiteV} we discuss generic features of finite-volume effects on QFT.
Although we consider degenerate vaccua, the introduction of a source $j$ to define the 1PI effective theory
lifts this degeneracy, which in principle leads to the formation of bubbles of different vacuua, as described
originally in \cite{CC}. 
The radius of such a bubble goes to infinity for a vanishing source though, which is not consistent with a finite volume
if we focus on the  
vicinity of the true vacuum $\phi=0$, which is mapped to $j=0$ through the Legendre transform. 
Hence bubbles cannot form and, given the $O(4)$
symmetry between Euclidean spacetime components, only homogeneous saddle points play a role in the partition 
function.

We calculate the one-loop 1PI effective potential in Sec.\ref{Ueff}, starting from the semi-classical approximation 
for the partition function, based on the two homogeneous saddle points. An important technical point consists in 
renormalising the connected generating functionals for each individual saddle point first, before 
performing the Legendre transform. As expected, the effective potential obtained from the interplay of 
the two saddle points is convex. Another important feature is a non-trivial spacetime volume dependence, 
such that the effective action is not extensive.

Sec.\ref{energetics} describes the consequences of the above features, in particular the NEC violation. 
This violation is a dynamical processs, arising from quantum fluctuations, and is a consequence of 
finite spacetime volume.

\section{Finite volume effects}\label{finiteV}

We consider a single real scalar field in a double-well potential, described by the action  
\be
S[\phi]=\int d^4x~\frac{1}{2}\partial_\mu\phi\partial^\mu\phi+\frac{\lambda}{24}(\phi^2-v^2)^2+j\phi~,
\ee
featuring two vacuua which are degenerate when the source $j$ vanishes.

\subsection{Saddle points}

Tunnelling between the two vacuua should be taken into account in the situation where fluctuations above these overlap \cite{ACP},
which, assuming an $O$(4)-symmetric Euclidean space-time, is controled by the dimensionless parameter
\be\label{A}
A\equiv\frac{\lambda L^4 v^4}{24\hbar}~.
\ee
Tunnelling happens for any finite volume and thus any finite parameter $A$, but with a probability decreasing exponentially with $A$. 
For this reason, the larger $A$, the longer one should wait for the equilibrium to be reached.
\\

Besides homogeneous solutions of the equation of motion, 
one should in principle consider other saddle points, which depend on the the four-dimensional Euclidean radial coordinate $r=\sqrt{x^\mu x^\mu}$.
We impose periodic boundary conditions in the finite four-dimensional volume where the scalar field is, so that a shot saddle point \cite{Andreassen} cannot 
be taken into account. For a non-vanishing source $j\ne0$, one could in principle have bubbles of different vacuua, but
we explain here why we can disregard these bubbles. 

The radius of the four-dimensional bubble is obtained by minimising the bubble action,
which corresponds to a compromise between the energy inside the bubble and the surface tension at the bubble wall \cite{CC}. This leads to
\be
R\simeq\frac{3v^2}{2j}\sqrt\frac{3}{\lambda}~,
\ee
and should be smaller than $L$, which leads to the minimum value for the source 
\be\label{boundj}
j\ge\frac{3v^2}{L}\sqrt\frac{3}{\lambda}~.
\ee
Because of symmetry restoration via tunnelling, we focus on the true vacuum $\phi_c=0$, which maps to $j=0$ through the Legendre transform 
used in the derivation of the 1PI effective action. Thus we restrict the present study to sources smaller than the lower bound (\ref{boundj}),
for which bubble saddle points cannot form. In what follows we therefore consider homogeneous saddle points only, which 
satisfy periodic boundary conditions. 

The equation of motion for these saddle points is
\be\label{equasaddle}
\phi^3-v^2\phi+\frac{6j}{\lambda}=0~,
\ee
with a number of real solutions depending on the source. We introduce the 
critical source $j_c\equiv\lambda v^3/(9\sqrt3)$, as well as the dimensionless source $k\equiv j/j_c$:
\begin{itemize}
\item If $|k|>1$, there is one (real) solution, which is
\be\label{varphi0s}
\varphi_0=-\mbox{sign}(k)\frac{2}{\sqrt3}\cosh\left(\frac{1}{3}\cosh^{-1}(|k|)\right)~.
\ee
This corresponds to the regime where the partition function is defined above one saddle point, 
and the 1PI effective potential has the known expression (\ref{Ueffext}) given further in this article;
\item If $|k|<1$ there are two solutions, which are 
\bea
\phi_1(k)&=&\frac{2v}{\sqrt3}\cos\Big(\pi/3-(1/3)\arccos(k)\Big)\\
\phi_2(k)&=&\frac{2v}{\sqrt3}\cos\Big(\pi-(1/3)\arccos(k)\Big)\nn
&=&-\phi_1(-k)~,\nonumber
\eea
and is the regime this article focuses on. We note that, in the limit $k\to 1^-$, the semi-classical 
approximation for the partition function might not be accurate, since fluctuations above the "false vacuum" (the one with the highest energy) are large.
In this article we restrict our studies to values $k<<1$ though, where the semi-classical approximation is reliable.
\end{itemize}

\subsection{Discrete versus continuous momentum components}

Quantum corrections about a uniform saddle point $\phi_i$ are in principle quantised in a finite box, and with periodic boundary conditions the
corresponding fluctuation determinant is
\bea
&&\frac{1}{\sqrt{\mbox{det}(\delta^2S[\phi_i])}}\\
&=&\exp\left(-\frac{1}{2}\sum_{n_\mu}
\ln\left(\frac{(2\pi/L)^2 n_\mu n_\mu+U''(\phi_i)}
{(2\pi/L)^2 n_\mu n_\mu+\lambda v^2/3} \right)\right)~,\nonumber
\eea
where $n_\mu$ is a set of integers, each going from $-\infty$ to $\infty$.
In what follows we will approximate the sum over 
discrete wave vector components $2\pi n_\mu/L$ by an integral over continuous components. 
To justify this, we focus for simplicity on the summation over the index $n_0$, and we introduce the functions
\bea
F(\Phi)&=&\sum_{n_0=-\infty}^\infty \ln(n_0^2+\Phi^2)\\
G(\Phi)&=&\int_{-\infty}^\infty dx~\ln(x^2+\Phi^2)~,\nonumber
\eea
where $\Phi^2=n_1^2+n_2^2+n_3^2+(L/2\pi)^2U''(\phi_i)$. The derivatives of these functions are easy to calculate
\bea
F'(\Phi)&=&2\Phi\sum_{n_0=-\infty}^\infty\frac{1}{n^2+\Phi^2}=2\pi\coth(\pi\Phi)\\
G'(\Phi)&=&2\Phi\int_{-\infty}^\infty\frac{dx}{x^2+\Phi^2}=2\pi~,\nonumber
\eea
and one can see that they are approximately identical for $\Phi\gg1/\pi$, which is indeed the case if the integers $n_1,n_2,n_3$ 
are not small. Quantum corrections are dominated by large integers $n_1,n_2,n_3$ though, and are therefore well approximated by continuous summation over
wave vector components.

\section{Interplay of saddle points}\label{Ueff}

In this section we derive the one-loop effective action based on two homogeneous saddle points, extending the work \cite{AT} in the situation of one
scalar flavour. 

\subsection{Semi-classical approximation}

The semi-classical approximation for the partition function evaluated from the two uniform saddle points assumes that different saddle points are 
"far enough" in field configuration space, for the quadratic fluctuations above these saddle points to be independent
\bea\label{semiclassical}
Z[k]&\simeq&\sum_{i=1,2}\frac{\exp(-S[\phi_i]/\hbar)}{\sqrt{\mbox{det}(\delta^2 S[\phi_i])}}\\
&=&\sum_{i=1,2}\exp(-\Sigma[\phi_i]/\hbar)~,\nonumber
\eea
where the individual connected-graph generating functionals are
\bea
\Sigma[\phi_i]&\equiv& S[\phi_i]+\frac{\hbar}{2}\mbox{Tr}\left\{\ln\left(\delta^2 S[\phi_i]\right)\right\}\\
&=&S[\phi_i]+\hbar\frac{V}{2}\int\frac{d^4p}{(2\pi)^4}
\ln\left(\frac{p^2+U''(\phi_i(k))}{p^2+U''(\phi_i(0))}\right)\nn
&=&Vj\phi_i+\frac{\lambda V}{24}(\phi_i^2-v^2)^2\nn
&&+\frac{\hbar\lambda^2v^4}{288\pi^2}V\int_0^X xdx\ln\left(\frac{x+a_i}{x+1}\right)~.\nonumber
\eea
In the latter expression, 
\be
X=\frac{3\Lambda^2}{\lambda v^2}~~~~\mbox{and}~~~~a_i=\frac{3}{2}\frac{\phi_i^2}{v^2}-\frac{1}{2}~,
\ee
and $\Lambda$ is an ultraviolet cut off.

\subsection{Renormalised  connected graphs generating functional}

Instead of renormalising the 1PI effective action $\Gamma$, we first renormalise the individual connected-graph generating functionals $\Sigma[\phi_i]$ 
and then perform the Legendre transform to find the renormalised effective action $\Gamma$. 
This procedure avoids a potential confusion arising from mixing the different loop orders from both $\Sigma[\phi_i]$.
Keeping the dominant terms in $\Lambda$ in the above loop integral, we obtain
\bea
&&\frac{1}{V}\Sigma[\phi_i]=j\phi_i\\
&&-\frac{\lambda v^2}{12}\left(1-\frac{\hbar}{16\pi^2}\left(3\frac{\Lambda^2}{v^2}+\frac{1}{2}\lambda\ln\left(\frac{\Lambda^2}{\lambda v^2} \right)\right)\right)\phi_i^2\nn
&&+\frac{\lambda}{24}\left(1-\frac{3\hbar\lambda}{32\pi^2}\ln\left(\frac{\Lambda^2}{\lambda v^2}\right)\right)\phi_i^4\nn
&&+\frac{\hbar\lambda^2v^4}{2304\pi^2}\left(3\frac{\phi_i^2}{v^2}-1\right)^2\ln\left(\frac{3}{2}\frac{\phi_i^2}{v^2}-\frac{1}{2}\right)~,\nonumber
\eea
where source-independent terms are omitted, as well as terms which vanish in the limit $\Lambda\to\infty$. 
We then define the renormalised parameters 
\bea
\lambda_R&\equiv&\lambda-\frac{3\hbar\lambda^2}{32\pi^2}\ln\left(\frac{\Lambda^2}{\lambda v^2}\right)\\
v_R^2&=&v^2-\frac{\hbar v^2}{16\pi^2}\left(3\frac{\Lambda^2}{v^2}-\lambda \ln\left(\frac{\Lambda^2}{\lambda v^2}\right)\right)~,\nonumber
\eea
and we obtain 
\be
\Sigma[\phi_i]=\Sigma_R[\phi_i]+{\cal O}(\hbar^2)~,
\ee
where
\bea\label{Sigma}
\frac{1}{V}\Sigma_R[\phi_i]&=&j\phi_i-\frac{\lambda_Rv_R^2}{12}\phi_i^2+\frac{\lambda_R}{24}\phi_i^4\\
&+& \frac{\hbar\lambda_R^2v_R^4}{2304\pi^2}\left(3\frac{\phi_i^2}{v_R^2}-1\right)^2\ln\left(\frac{3}{2}\frac{\phi_i^2}{v_R^2}-\frac{1}{2}\right)~.\nonumber
\eea
In the latter expression, the saddle points $\phi_i$ are still expressed in terms of the bare parameters $\lambda$ and $v$. But one can write, in terms of the 
renormalised dimensionless source $k_R=9\sqrt3j/(\lambda_Rv_R^3)$ and the renormalised saddle point $\phi_{iR}(k_R)=\phi_i(k)+{\cal O}(\hbar)$,
\bea
\Sigma_R[\phi_i(k)]&=&\Sigma_R[\phi_{iR}(k_R)]\\
&+&(\phi_i(k)-\phi_{iR}(k_R))\left.\frac{\partial\Sigma_R}{\partial\phi_i}\right|_{\phi_{iR}(k_R)}+{\cal O}(\hbar^2)\nn
&=&\Sigma_R[\phi_{iR}(k_R)]\nn
&+&(\phi_i(k)-\phi_{iR}(k_R))\left.\frac{\partial\Sigma}{\partial\phi_i}\right|_{\phi_i(k)}+{\cal O}(\hbar^2)~.\nonumber
\eea
Since the equation of motion satisfied by the saddle points are $\partial\Sigma/\partial\phi_i={\cal O}(\hbar)$, we finally obtain
\be
\Sigma_R[\phi_i(k)]=\Sigma_R[\phi_{iR}(k_R)]+{\cal O}(\hbar)^2~,
\ee
and the saddle points in the expression (\ref{Sigma}) can be read as functions of $\lambda_R$ and $v_R$.
We introduce the dimensionless quantities
\be
\varphi_i\equiv\frac{\phi_{iR}(k_R)}{v_R}~~~~\mbox{and}~~~~A_R\equiv\frac{\lambda_RVv_R^4}{24\hbar}~,
\ee
in terms of which we write the final expression for the renormalised individual connected graphs generating functionals as
\bea\label{SigmaR}
\Sigma_R[\varphi_i]&=&\hbar A_R\Bigg(\frac{8k_R}{3\sqrt3}\varphi_i-2\varphi_i^2+\varphi_i^4\\
&+&\frac{\hbar\lambda_R}{96\pi^2}(3\varphi_i^2-1)^2\ln\left(\frac{3}{2}\varphi_i^2-\frac{1}{2}\right)\Bigg)~,\nonumber
\eea
where 
\bea
\varphi_1(k_R)&=&\frac{2}{\sqrt3}\cos\Big(\pi/3-(1/3)\arccos(k_R)\Big)\\
\varphi_2(k_R)&=&\frac{2}{\sqrt3}\cos\Big(\pi-(1/3)\arccos(k_R)\Big)\nn
&=&-\varphi_1(-k_R)~.\nonumber
\eea
Below we derive the effective potential obtained from the two functionals (\ref{SigmaR}) in the semi-classical approximation (\ref{semiclassical}).

\subsection{One-loop 1PI effective action}\label{One-loop 1PI effective action}

We follow here the usual steps leading to the 1PI effective action.
Starting from the partition function (\ref{semiclassical}), the classical field is given by
\be\label{Classical Field}
\frac{\phi_c}{v_R}=-\frac{3\sqrt3}{8A_R}\frac{\partial\ln(Z)}{\partial k_R}~,
\ee
and an expansion in the source $k_R$ gives
\be
\frac{\phi_c}{v_R}=f_1k_R+f_3k_R^3+\Od(k_R^5)~,
\ee
where, at one-loop,
\bea
f_1&=&-\frac{1+8A_R}{3\sqrt{3}}
+\frac{\hbar\lambda_R}{192\sqrt{3}\pi^2}\Big(7 + 16 A_R\Big)\\
f_3&=&\frac{4}{243\sqrt{3}}\Big(128A_R^3+12A_R-3\Big)\nn
&&-\frac{\hbar\lambda_R}{15552\sqrt{3}\pi^2}\Big(2048A_R^3+384A_R-93\Big)~.\nonumber
\eea
In order to perform the Legendre transform, we inverted the latter relation as
\bea
k_R&=&-\frac{3\sqrt3}{8}g_2\left(\frac{\phi_c}{v_R}\right)-\frac{\sqrt3}{16}g_4\left(\frac{\phi_c}{v_R}\right)^3\\
&&+\Od(\phi_c/v_R)^5~,\nonumber
\eea
where the one-loop coefficients are
\bea\label{g}
g_2&=& \frac{8}{1+8A_R}
+\frac{\hbar\lambda_R}{8\pi^2}~\frac{16A_R+7}{(1+8A_R)^2}\\
g_4&=&\frac{64(128A_R^3+12A_R-3)}{(1+8A_R)^4}\nn
&&+\frac{\hbar\lambda_R}{4\pi^2}~\frac{16384 A_R^4+12288 A_R^3+936 A_R-243}{(1+8A_R)^5}~.\nonumber
\eea
The renormalised effective action for a constant classical field $\Gamma=VU_{eff}$ is obtained by integrating the equation
\be
\frac{dU_{eff}}{d\phi_c}=-j ~~~~\rightarrow~~~~  \frac{d\Gamma}{d\phi_c}=-\frac{8\hbar A_R}{3\sqrt3}\frac{k_R}{v_R}~,
\ee
such that finally
\bea\label{Gamma}
\Gamma[\phi_c]&=&\hbar A_R\Bigg(g_0+\frac{g_2}{2}\left(\frac{\phi_c}{v_R}\right)^2
+\frac{g_4}{24}\left(\frac{\phi_c}{v_R}\right)^4\\
&&~~~~~~~+\Od(\phi_c/v_R)^6\Bigg)~,\nonumber
\eea
where $g_0$ is a constant. We note that the limit $\hbar\to0$ leads to the result derived in \cite{AT}, where the effective action is 
derived in the "tree-level semi-classical approximation", i.e. with fluctuation factors which are ignored.  
The one-loop effective action satisfies the same fundamental features as the one obtained in \cite{AT}:
\begin{itemize}
    \item Consistently with general arguments, the action (\ref{Gamma}) is convex, since $g_2>0$ and $g_4>0$ for large $A_R$;
    \item The action (\ref{Gamma}) is not extensive, as a result of the non-trivial volume dependence of the renormalised constants $g_2,g_4$,
    through the parameter $A_R$.
\end{itemize}

\subsection{Resummation for infinitesimal source}\label{Resummation for infinitesimal source}

It is interesting to note that, in the limit of infinitesimal source, one can obtain an analytical expression for the one-loop effective
action. If we consider terms linear in $k_R$ only in the renormalised connected graphs generating functional, we obtain
\be
\Sigma_R[\varphi_1(k_R)]=\frac{8\hbar A_R}{3\sqrt3~\eta}~ k_R
=-\Sigma_R[\varphi_2(k_R)]~,
\label{1PI Large A}
\ee
where source-independent terms are dropped, and
\be
\frac{1}{\eta}\equiv1-\frac{\hbar\lambda_R}{64\pi^2}~.
\ee
The steps described above can be applied to eq.\eqref{1PI Large A} without the need for an expansion in $k_R$, and they lead to
the classical field 
\be
\eta\frac{\phi_c}{v_R}=-\tanh\left(\frac{8 A_R}{3\sqrt3~\eta}~ k_R\right)~.
\ee
This relation is inverted as
\be
\frac{8 A_R}{3\sqrt3} k_R=\frac{\eta}{2}\ln\left(\frac{1-\eta\phi_c/v_R}{1+\eta\phi_c/v_R}\right)~,
\ee
and leads to the one-loop effective action
\bea\label{resum}
\tilde\Gamma[\phi_c]&=&\hbar A_Rg_0
+\frac{\hbar}{2}\left(1-\eta\frac{\phi_c}{v_R}\right)\ln\left(1-\eta\frac{\phi_c}{v_R}\right)\\
&&+\frac{\hbar}{2}\left(1+\eta\frac{\phi_c}{v_R}\right)\ln\left(1+\eta\frac{\phi_c}{v_R}\right)~,\nonumber
\eea
where the integration constant is chosen to match the effective action \eqref{Gamma}. This result was derived in \cite{A} in the "tree-level semi-classical approximation", and the one-loop result (\ref{resum}) 
simply consists in the replacement $v\to v_R/\eta$, which leaves the functional form of the effective action unchanged. 
The expression (\ref{resum}) apparently makes sense for $|\phi_c|\le v_R/\eta$ (although it is  
not differentiable at $|\phi_c|= v_R/\eta$), but because of the one-to-one mapping between $\phi_c$ and $k_R\ll1$, the effective action (\ref{resum}) 
is actually valid for $|\phi_c|\ll v_R/\eta$ only. 

\vspace{0.5cm}

As a side comment, we show here that the effective action (\ref{resum}) 
corresponds to a resummation of all the orders in $\phi_c$, in the limit of large (but finite) volume $A_R\gg1$. 
To see this, let us express the functional $\Sigma_R$ in terms of the original variables
\be
\Sigma_R=V\left(\epsilon jv_R +a\frac{j^2}{v_R^2}+\cdots\right)~,
\ee
where $\epsilon=\pm1$ (depending on which vacuum one focuses), $a$ is a dimensionless constant independent of $V$, and dots represent higher orders 
in $j$, which are also independent of $V$. Since we are interested in the vicinity of the vacuum $\phi_c=0$, 
it is enough to choose a source which satisfies $|Vjv_R|\ll1$. In this situation, the quadratic term in the action $\Sigma_R$ is of the order
\be
\frac{j^2}{v_R^2}V=|Vjv_R|~\frac{|j|}{v_R^3}\ll |Vjv_R|\frac{1}{Vv_R^4}~,
\ee
which is therefore negligible compared to the linear term $Vjv_R$ for large volume $V$. As a consequence, the limit $A_R\gg1$ of
the effective action (\ref{Gamma}) should be identical to the Taylor expansion of the resummation (\ref{resum}).
One can check that this is indeed the case, since
\bea
\Gamma[\phi_c]&=&
\hbar A_R g_0
+\frac{\hbar}{2}\left(1+\frac{\hbar\lambda_R}{32\pi^2}
+{\cal O}\left(A_R^{-1}\right)\right)\left(\frac{\phi_c}{v_R}\right)^2\nn
&&+\frac{\hbar}{24}\left(2+\frac{\hbar\lambda_R}{8\pi^2}
+{\cal O}\left(A_R^{-1}\right)\right)\left(\frac{\phi_c}{v_R}\right)^4\\
&&+{\cal O}(\phi_c/v_R)^6\nn
&=&\hbar A_R g_0+\frac{\hbar}{2}\Big(1+{\cal O}\left(A_R^{-1}\right)\Big)\left(\eta\frac{\phi_c}{v_R}\right)^2\nn
&&+\frac{\hbar}{12}\Big(1+{\cal O}\left(A_R^{-1}\right)\Big)\left(\eta\frac{\phi_c}{v_R}\right)^4\nn
&&+{\cal O}(\phi_c/v_R)^6~,\nonumber
\eea
where higher orders in $\hbar$ are neglected. On the other hand, we have
\bea
\tilde\Gamma[\phi_c]&=&\hbar A_Rg_0
+\frac{\hbar}{2}\left(\eta\frac{\phi_c}{v_R}\right)^2
+\frac{\hbar}{12}\left(\eta\frac{\phi_c}{v_R}\right)^4\\
&&+{\cal O}(\phi_c/v_R)^6~,\nonumber
\eea
such that
$\Gamma[\phi_c]$ and $\tilde\Gamma[\phi_c]$ are identical up to terms of order $A_R^{-1}$, and the convergence as $A_R$ increases 
can be seen on Figs. \ref{Resummation Plots}. We also note that this identification is valid to all orders in $\phi_c$, although we 
show here the property up to the order 4 only.
We stress here that deriving the effective potential for higher powers of the classical field requires higher orders in the source, and thus
should take into account bubble saddle points, which is not done here. So the expression (\ref{resum}) is actually valid for $|\phi_c|\ll v_R$,
although a similar resummation might exist in the presence of bubbles, which is a topic to be explored in a future work.

\section{Energetics}\label{energetics}

We discuss in this section the energetic properties arising form the non-extensive feature of the effective action (\ref{Gamma}).

\subsection{Matching with the single-saddle-point regime}

The effective potential for $|k_R|>1$ is based on a single saddle point, and has the usual expression in terms of the renormalised parameters $\lambda_R,v_R$
\bea\label{Ueffext}
&&U_{eff}^{(|k_R|>1)}(\phi_c)
=\frac{\lambda_R}{24}(\phi_c^2-v_R^2)^2\\
&&+\frac{\hbar\lambda_R^2}{2304\pi^2}(3\phi_c^2-v_R^2)^2\ln\left(\frac{3}{2}\frac{\phi_c^2}{v_R^2}-\frac{1}{2}\right)~,
\nonumber
\eea
where the origin of energies is chosen such that $U_{eff}^{(|k_R|>1)}(v_R)=0$.

\begin{figure}
    \centering
    \begin{subfigure}{0.8\linewidth}
        \centering
        \includegraphics[width=\linewidth]{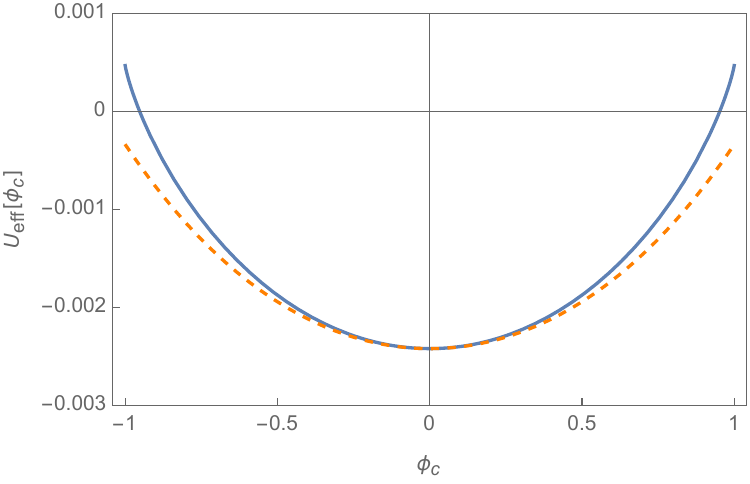}
        \caption{$A_R=1$}
        \label{1}
    \end{subfigure}
     \hfill
     \begin{subfigure}{0.8\linewidth}
         \centering
         \includegraphics[width=\linewidth]{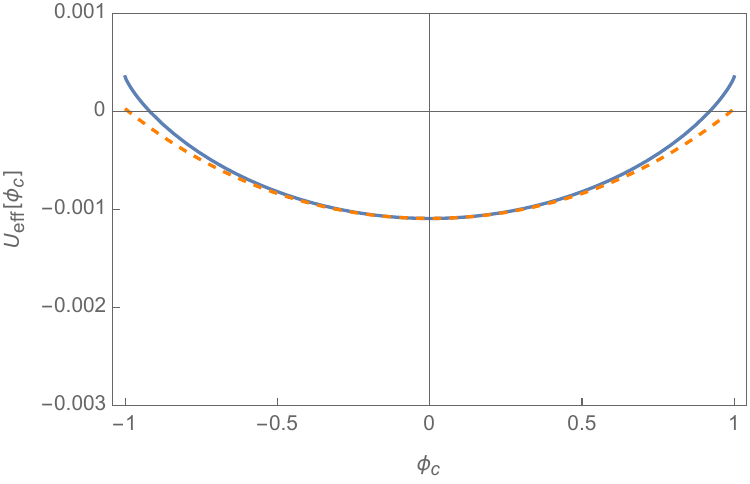}
         \caption{$A_R=2$}
         \label{5}
     \end{subfigure}
     \hfill
     \begin{subfigure}{0.8\linewidth}
         \centering
         \includegraphics[width=\linewidth]{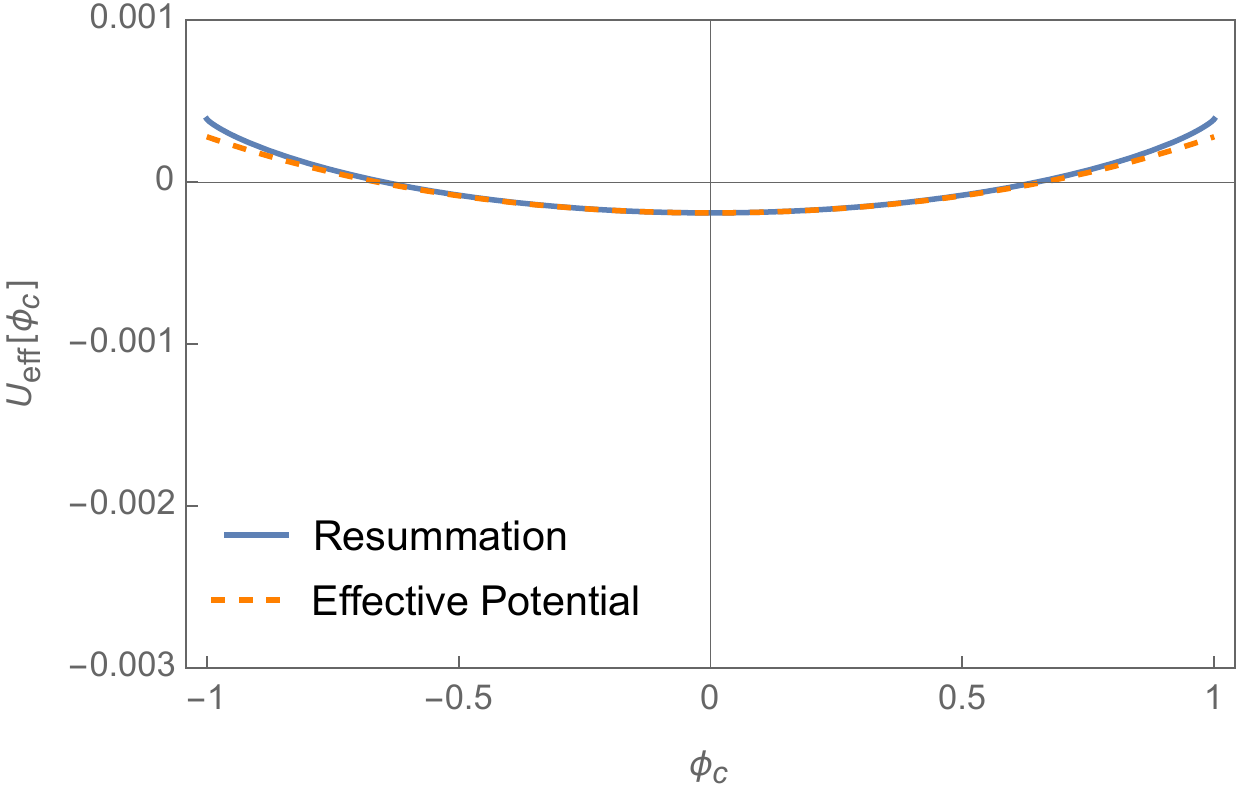}
         \caption{$A_R=5$}
         \label{10}
     \end{subfigure}
    \caption{Plots of the effective potential and the resummation for $\lambda=0.1$, $\hbar=v_R=1$, and a range of values for $A_R$.}
    \label{Resummation Plots}
\end{figure}

\begin{figure}
    \centering
    \includegraphics[width=\linewidth]{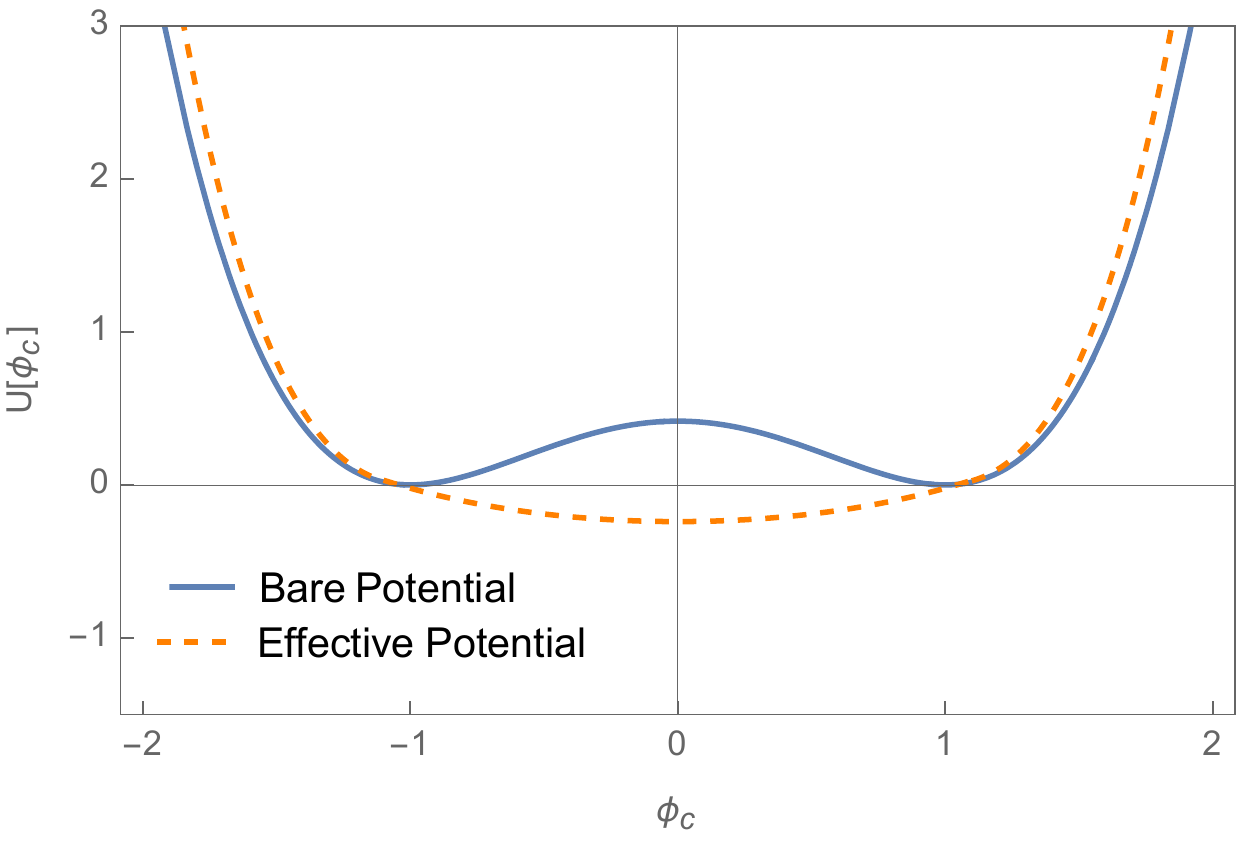}
    \caption{Plots of the bare and effective potentials for $\hbar=A_R=v_R=1$ and $\lambda_R=10$. 
    This choice of $\lambda_R$ corresponds to a strong coupling regime, but is chosen for demonstrative purposes.}
    \label{Effective Potential}
\end{figure}

The constant $g_0$ in eq.(\ref{Gamma}) is obtained by imposing continuity of the effective potential at $|k_R|=1$, 
which corresponds to the boundary between the regime with one saddle point and the regime with two homogeneous saddle points.
Taking the limit of eq.\eqref{Classical Field} for $|k_R|\to1$ yields the corresponding classical field
\be
\phi_c(|k_R|\to1)=\pm\frac{v_R}{\sqrt{3}}~\frac{1-2e^{3A_R}}{1+e^{3A_R}}~,
\ee
which, for large $A_R$, takes the value $\phi_c(|k_R|\to1)\simeq\pm2v_R/\sqrt3$. The requirement 
\be
U_{eff}^{(|k_R|>1)}(2v_R/\sqrt3)=U_{eff}^{(|k_R|<1)}(2v_R/\sqrt3)~,
\ee
leads then to
\be
g_0\simeq \frac{1}{9}-\frac{2}{3}g_2-\frac{2}{27}g_4+\frac{3\hbar\lambda_R}{32\pi^2}\ln\left(\frac{3}{2}\right)~,
\ee
and this expression is used below to compare the effective potential with the bare potential 
(see Fig. \ref{Effective Potential}), 
and with the resummation \eqref{resum} for different values of $A_R$ (see Fig. \ref{Resummation Plots}).

\subsection{Null Energy Condition}

\begin{figure}
    \centering
    \includegraphics[width=\linewidth]{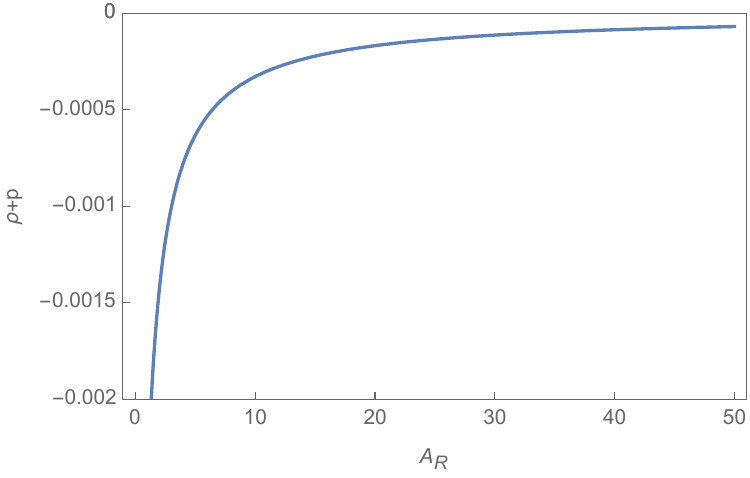}
    \caption{A plot of $\rho+p$ as a function of $A_R$ for $\lambda=0.1$ and $\hbar=v_R=1$: the NEC is violated for a finite spacetime volume $V$,
    and is recovered asymptotically when $V\to\infty$.}
    \label{NECPlot}
\end{figure}

The fact that all known matter satisfies the NEC is an important conjecture, 
since it appears as one of the assumptions for the derivation of singularity theorems \cite{HP}.
For a homogeneous fluid with density $\rho$ and pressure $p$, the NEC reads $\rho+p\ge0$,
and the fluid we consider here is the ground state $\phi_c=0$.
Although a full study of the mechanism presented here should be extended to curved spacetime,
we can already see with the result (\ref{Gamma}) how tunnelling can lead to a dynamical NEC violation, 
as a consequence of the 
non-extensive nature of the effective theory \cite{ACP}. Following the thermodynamical approach, we have
\bea\label{rho+p}
&&\rho+p\\
&=&\frac{\Gamma[0]}{V}-\frac{\partial\Gamma[0]}{\partial V}\nn
&=&-\frac{\lambda_Rv_R^4}{24}~V\frac{\partial g_0}{\partial V}\nn
&=&-\frac{A_R\lambda_R v_R^4}{81(1+8A_R)^5}\left(P_1(A_R)+\frac{\hbar\lambda_R}{\pi^2}\frac{P_2(A_R)}{1+8A_R}\right)~.\nonumber
\eea
with 
\bea
P_1(x)&=&16(5632x^3+1344x^2+504x-99)\\
P_2(x)&=&26624x^4+28928x^3+3744x^2+2556x-639~,\nonumber
\eea
and one can see that $\rho+p<0$. 
Also, this NEC violation is a finite volume effect, since for $A_R>>1$ we have
\bea\label{rho+plargeA}
\rho+p&\simeq&-\frac{\lambda_Rv_R^4}{324A_R}\left(11+\frac{13\hbar\lambda_R}{32\pi^2}\right)\\
&=&-\frac{22\hbar}{27V}+{\cal O}(\hbar^2)~.\nonumber
\eea
The result (\ref{rho+p}) is sketched as a function of $A_R$ in Fig. \ref{NECPlot}, and we finish this section with two important comments:
\begin{itemize}
    \item From the expression (\ref{rho+p}) and (\ref{rho+plargeA}), one can see that the NEC is violated for all finite values of $A_R$, 
    although we find numerically that the constant $g_0$ becomes positive for $A_R\gtrsim7$ when $\lambda=0.1$ and $\hbar=v_R=1$. 
    NEC violation is therefore independent of the origin of energies, which is expected in flat spacetime;
    \item Although the NEC is violated, the Averaged NEC (ANEC) is not. The latter is a weaker condition than the former, and requires instead
    \be
    \int d\lambda (\rho+p)\ge0~,
    \ee
    where the integral runs along a null geodesic. As explained in \cite{ANEC} with specific examples related to the Casimir effect or a more 
    generic confining potential, the environmental energy needed to maintain the scalar field confined compensates the negative value of $\rho+p$ 
    inside the box, leading to the ANEC being satisfied. 
\end{itemize}

\section{Conclusion}

We generalised the results of \cite{AT} and \cite{ACP} by taking into account one-loop corrections in the semi-classical approximation for the partition function. 
A consistent renormalisation of the model requires to redefine parameters before implementing the Legendre transform, which is specific to the presence of several 
saddle points. Our results show that the present NEC violation mechanism is stable under quantum fluctuations, and is a fundamental feature due to the structure of 
the partition function, independently of the accuracy of the latter.

The next steps we plan to make include:\\ 
{\it (i)} Large volume $L^4$: although tunnelling is suppressed, if one waits a long enough (real) time for the equilibrium to be reached,
the present mechanism should hold. In this case one should take into account additional saddle points for a non-vanishing source though, 
in the form of bubbles with different vacuua \cite{CC}. Among the next studies is to evaluate the contribution 
of the NEC violation effect described here to Dark Energy.\\ 
{\it(ii)} Curved spacetime: the $O(4)$ symmetry between space time coordinates is not valid if one 
focuses on a Friedman-Lemaitre-Robertson-Walker metric for example, and the finite-temperature study \cite{AP2} needs to be extended beyond flat spacetime. 
An appropriate causal quantisation volume should be defined, as well as a comparison between the tunnelling rate and the spacetime expansion rate. One should also include the 
non-extensive feature of the effective action in the energy-momentum tensor of the fluid to be coupled to (classical) gravity;\\ 
{\it(iii)} Real-time tunnelling: ideally these studies should be done in a Minkowski (or Lorentzian) metric. 
Tunnelling in real time is more involved though (see \cite{Ai} for a review), but it would allow to go beyond equilibrium field theory, which could be game-changing 
to study the Early Universe.

\section*{Acknowledgements}

The work of JA is supported by the Leverhulme Trust, grant RPG-2021-299, and the work of DB is supported by the STFC, grant ST/T000759/1.

\end{document}